\documentclass[twoside,twocolumn,tightenlines, showpacs,aps, prb]{revtex4}

\newcommand{\be}{\begin{equation}}  
\newcommand{\ee}{\end{equation}}

\begin{document}
\title{Diagrammatic Analysis of
the Unitary Group for Double Barrier \\
Ballistic Cavities:
Equivalence with Circuit Theory}

\author{A. L. R. Barbosa and  A. M. S. Mac\^edo}

\affiliation{Departamento de F\'{\i}sica, 
Laborat\'orio de F\'{\i}sica Te\'orica e Computacional,\\ 
Universidade Federal de Pernambuco
50670-901 Recife, PE, Brazil}

\begin{abstract}
We derive a set of coupled non-linear algebraic equations for the 
asymptotics of the Poisson kernel 
distribution describing the statistical properties of a two-terminal
double-barrier chaotic billiard (or ballistic quantum dot). The equations 
are calculated from a diagrammatic
technique for performing averages over the unitary group,
proposed by Brouwer and Beenakker [J. Math. Phys. {\bf 37}, 4904 (1996)].
We give strong analytical evidences that these equations are equivalent 
to a much simpler polynomial equation calculated from
a recent extension of Nazarov's
circuit theory [A. M. S. Mac\^edo, Phys. Rev. B {\bf 66}, 033306 (2002)]. 
These results offer interesting perspectives for further developments 
in the field via the direct conversion of one approach into the other.
\end{abstract}

\maketitle

\section{Introduction}

Random-matrix theory has proved to be a powerful tool for describing generic 
features of quantum chaotic systems \cite{ggw98,m91}. 
For closed systems it offers an accurate statistical description 
of both energy levels and wavefunctions characteristics with overwhelming numerical
and experimental confirmation. In this case, the central hypothesis is to replace the Hamiltonian of the system by a matrix with random independent gaussian entries 
obeying certain fixed exact symmetries, such as time-reversal and spin-rotation.
The resulting ensembles are known in the literature as the Gaussian Ensembles.
For open systems, such as ballistic chaotic cavities and disordered wires
attached to external leads, the major 
role is shifted from the Hamiltonian to the appropriate matrix associated with 
the description of electron transport, viz. 
the scattering matrix or the transfer 
matrix. Here, the justification of the specific ensemble is a subtle procedure, because,
unlike the previous case, one needs to take into account subtle correlations between the 
entries of the matrix induced by flux conservation and quantum diffusion (for disordered wires). 
\par
For ballistic
chaotic cavities, a general procedure to obtain such an ensemble was developed by
Mello and co-workers \cite{mb99}. 
It consists of evoking the {\it maximum information entropy
principle} along with certain generic assumptions, such as analyticity, unitarity
and some specific symmetries  that are exactly preserved in the presence of chaotic
dynamics. The resulting $S$-matrix distribution, $P(S)$, turns
out to be a multidimensional generalization of the Poisson kernel, usually found in 
two-dimensional electrostatics. Its general form reads

\begin{equation}
P(S)\propto\vert \det(1-\bar{S}^{\dagger}S)\vert^{-(\beta M+2-\beta)},
\label{nucleo}
\end{equation} 
where $\beta \in \{1,2,4\}$ is a parameter identifying distinct symmetry classes 
\cite{comment},
$M=2N$ is the total number of open scattering channels ($N$ for each waveguide) 
and $\bar{S}$ is a sub-unitary
matrix, i.e. the eigenvalues of $\bar{S}\bar{S}^{\dagger}$ may be less than unity.
For a two-terminal system, consisting of a ballistic chaotic cavity or quantum dot
coupled to two waveguides, the scattering matrix can be conveniently written as 
\cite{mb99}
\be
S=\left( 
\begin{array}{ll}
r & t \\ 
t^{\prime } & r^{\prime }
\end{array}
\right),
\label{Smatrix} 
\ee
where $t,t^{\prime}$ and $r,r^{\prime}$ are respectively transmission 
and reflection matrices. The presence of barriers of arbitrary transparencies
at the interface between the chaotic cavity and the waveguides can be accounted 
for by specifying the average (or optical part) of the scattering matrix 
\cite{ggw98,mb99}.
\begin{equation}
\bar{S}= \left( \begin{array}{cc}
r_1 & 0\\ 0 & r_2\\
\end{array} \right),
\label{Sm}
\end{equation} 
where  $r_1$ e $r_2$ are reflection matrices for barriers 1 and 2 
respectively.
\par
In mesoscopic physics, one is usually concerned with the description of 
a well defined measurement, such as the full counting statistics (FCS) 
of a two-terminal system, whose cumulant generating function is given 
by the Levitov-Lesovik formula \cite{ll93}.
\be
\Phi(\lambda)=-M_0 {\rm Tr}\ln(1+(e^{i\lambda}-1)tt^{\dagger}),
\label{Lev}
\ee
where $M_0=eVT_0/h \gg 1$ is the number of attempts to transmit an electron 
during the
observation time $T_0$, $t$ is the transmission matrix and $V$ is the voltage. 
Physical observables can be obtained from the series expansion
\be
\Phi (\lambda )=-\sum_{k=1}^\infty \frac{(i\lambda )^k}{k!}q_k,
\ee
where
\be
q_k=-\frac{d^k}{d(i\lambda )^k}\Phi (\lambda )\Big\vert_{\lambda =0}
\ee
are the irreducible cumulants of the FCS. Some well known examples are
the dimensionless conductance and shot-noise power, which are given 
respectively by
\be
g=q_1/M_0={\rm Tr}(tt^{\dagger}),
\label{g}
\ee
which is the Landauer formula, and
\be
p=q_2/M_0={\rm Tr}[tt^{\dagger}(1-tt^{\dagger})].
\label{p}
\ee
The third cumulant has also attracted some recent interest, including
an experimental observation in tunnel junctions \cite{rsp03}. It is given by
\be
\kappa=q_3/M_0={\rm Tr}[tt^{\dagger}(1-tt^{\dagger})(1-2tt^{\dagger})].
\label{r}
\ee

The above expressions for the cumulants of the FCS are sample specific, 
which means for a chaotic cavity
that an average over the Poisson kernel distribution is
necessary in order to make comparisons with real experimental data.
Such calculations can in principle be performed exactly for arbitrary 
values of $M$ and $\beta$ using the method of supersymmetry \cite{e97,vwz85} and
some very general results are in fact already available in the literature
\cite{amsm01,amsm04}. 
Albeit powerful and general, this method can become very cumbersome
and sometimes unwieldy. 
Fortunately, if one is interested only in the semiclassical regime,
where one neglects quantum interference contributions, defined 
mathematically by the asymptotic condition $M \gg 1$, 
much simpler alternative techniques exist. 
\par
Two well-known methods to deal with the semiclassical regime are 
the diagrammatic analysis of the 
unitary group, presented by Brouwer and Beenakker \cite{bb96} and Nazarov's
circuit theory \cite{n95}. In its original form circuit theory does not allow
for a direct comparison with the diagrammatic approach in the entire range
of parameters, i.e. for arbitrary values of the barrier's transparencies,
because of the intrinsic difficulty to determine the average 
pseudo-current-voltage characteristics of an arbitrary 
circuit element. This is related to
the well known problem of breakdown of semiclassical transport equations
close to boundaries and interfaces, where pure quantum effects become
dominant.
This difficulty was recently removed by a novel systematic 
treatment presented in Ref. \onlinecite{amsm02}, 
in which circuit theory is combined with the 
supersymmetric non-linear sigma model
yielding a powerful technique with a very convenient algebraic structure. 
This extended version of circuit theory raises the natural question as
to whether its information content coincides with that of the diagrammatic 
method in all ranges of the input parameters. This is a highly non-trivial 
question, because the semiclassical 
concatenation principle, which is the basis of circuit theory, does not 
have a direct representation in the diagrammatic formulation.
In particular, there is no obvious way how to extend this 
concatenation principle to account for quantum corrections, such as
weak-localization.
We remark that the semiclassical concatenation principle is directly
related to the independence of the leading asymptotic contribution
on the symmetry parameter $\beta$. By contrast, the weak localization
correction is strongly $\beta$ dependent \cite{bb96} and, in particular, vanishes
for $\beta=2$, i.e. for systems with broken time-reversal symmetry.
Interestingly, for the particular case of symmetric barriers a 
direct comparison between both approaches
was presented in Ref. \onlinecite{amsm02} and complete agreement was found.
\par
Motivated by this result, we present in this work a detailed comparison between 
the diagrammatic approach 
and the extended version of circuit theory for the case of asymmetric barriers
with arbitrary transparencies,
thus completing the analysis of Ref. \onlinecite{amsm02}
and providing strong evidence for the
full equivalence between these semiclassical techniques.
This result is particularly relevant in practical applications because
of the great algebraic advantage of the calculations in circuit theory (in our
case study the problem is reduced to a polynomial equation of fourth order)
in comparison with alternative approaches.
In fact, we believe that circuit theory equations reach the ultimate
irreducible form in terms of simplifying the description. 
Furthermore, when combined with the scaling theory presented in Ref. 
\onlinecite{amsm02} for the 
balistic-diffusive crossover in phase-coherent metallic conductors, in
which the present results enters as an initial condition, we 
end up with a very powerful formalism for performing concrete calculations
in realistic conductors. Applications might include 
spintronics \cite{zfs04}
and normal-superconducting hybrid systems \cite{pc00}. 

\section{The Diagrammatic Technique}
\par
In this section we present a detailed account of the application of the
diagrammatic method to the calculation of the semiclassical limit of
a double barrier chaotic ballistic cavity. It contains the central
result of this paper and complements previous analysis by Brouwer 
and Beenakker \cite{bb96}.
\par
We start by noting that all cumulants of the FCS can be generically 
written in the form
\be
A={\rm Tr}[a(tt^{\dagger})],
\ee
where $a(x)$ is an arbitrary smooth function. It means that the 
ensemble averages of transport observables can be fully described by 
the following density function
\be
\rho(\tau)=\langle {\rm Tr}\delta(\tau-tt^{\dagger})\rangle,
\ee
so that
\be
\langle A \rangle=\int_0^1 d\tau a(\tau)\rho(\tau).
\ee
The problem is thus reduced to the calculation of $\rho(\tau)$ in 
the asymptotic regime $M=2N \gg 1$.
\par
In the diagrammatic formalism, one starts by introducing the 
generating functions
\be
F_{1}(z)=\langle C_1(z-S^{\dagger}C_2SC_1)^{-1}\rangle,
\label{F1}
\ee
and
\be
F_{2}(z)=\langle C_2(z-SC_1S^{\dagger}C_2)^{-1}\rangle,
\label{F2}
\ee
where $S$ is the $M\times M$ scattering matrix defined
in Eq. (\ref{Smatrix}) and the
auxiliary matrices $C_1$ and $C_2$ are defined as
\begin{equation}
C_{1}= \left( \begin{array}{cc}
0 & 0\\ 0 & {1}_{N}\\
\end{array} \right),
\qquad
C_{2}= \left( \begin{array}{cc}
{1}_{N} & 0\\ 0 & 0\\
\end{array} \right),
\label{C1C2}
\end{equation} 
in which ${1}_{N}$ is the $N\times N$ unit matrix.
It can be easily verified that
\begin{equation}
\rho(\tau)=-\frac{1}{\pi}{\rm Im}[f_{1}(\tau+i0^{+})]
=-\frac{1}{\pi}{\rm Im}[f_{2}(\tau+i0^{+})],
\label{dt}
\end{equation} 
where $f_i(z)\equiv{\rm Tr}(F_i(z))$, $i=1,2$. 
\par
The next step is to define the matrix generating function
\be
\hat{F}(z)= \left( \begin{array}{cc}
0 & F_1(z)\\ F_2(z) & 0\\
\end{array} \right),
\ee
which can be written as a sum of two terms that can
be averaged separately, $\hat{F}(z)=(2z)^{-1}
(\hat{F}_{+}(z)+\hat{F}_{-}(z))$. Each term is given by
\be
\hat{F}_{\sigma}(z)=\hat{C}+\sigma\hat{A}(\hat{1}-\hat{\Sigma}_{\sigma}
\hat{G}^{0}_{\sigma})^{-1}\hat{\Sigma}_{\sigma}\hat{B},
\label{Fs}
\ee
where $\hat{A}=\hat{C}\hat{L}$, $\hat{B}=\alpha\hat{T}\hat{C}$,
with $\alpha \equiv z^{-1/2}$ and the following matrices have been
defined
\be
\hat{C}= \left( \begin{array}{cc}
0 & C_{1}\\ C_{2} & 0\\
\end{array} \right),
\qquad
\hat{L}= \left( \begin{array}{cc}
L & 0\\ 0 & L^{\dagger}\\
\end{array} \right),
\ee
and
\be
\hat{T}= \left( \begin{array}{cc}
T & 0\\ 0 & T^{\dagger}\\
\end{array} \right),
\qquad
\hat{R}= \left( \begin{array}{cc}
R & 0\\ 0 & R^{\dagger}\\
\end{array} \right).
\ee
The submatrices $L$, $T$ and $R$ describe transmission and reflection coefficients of
the barriers and are related to $\bar{S}$ of Eq. (\ref{Sm}) via the condition that
\begin{equation}
\hat{U}= \left( \begin{array}{cc}
\bar{S} & L\\ T & R\\
\end{array} \right),
\end{equation} 
be unitary. The remaining matrices in Eq. (\ref{Fs}) are the
``free propagator''
\be
\hat{G}^{0}_{\sigma}=\hat{R}+\sigma\alpha\hat{T}\hat{C}\hat{L}
\label{G0}
\ee
and the  ``self-energy'' matrix, $\hat{\Sigma}_{\sigma}$. They satisfy
a Dyson equation
\be
\hat{G}_\sigma=\hat{G}^0_\sigma+\hat{G}^0_\sigma\hat{\Sigma}_\sigma\hat{G}_\sigma
=\hat{G}^0_\sigma+\hat{G}_\sigma\hat{\Sigma}_\sigma\hat{G}^0_\sigma.
\label{Dyson}
\ee
Summation over planar diagrams \cite{bb96} yields the following 
expression for the self-energy matrix
\be
\hat{\Sigma}_{\sigma}=\hat{P}_\sigma \zeta(\hat{P}^2_\sigma),
\label{se}
\ee
where
\be
\hat{P}_\sigma =\left( 
\begin{array}{ll}
0 & {\rm Tr}(G_\sigma ^{12}) \\ 
{\rm Tr}(G_\sigma ^{21}) & 0
\end{array}
\right) \otimes 1_M,
\label{Proj}
\ee
in which $G_\sigma ^{12}$ and $G_\sigma ^{21}$
are off-diagonal blocks of the full Green´s
function
\be
\hat{G}_\sigma =\left( 
\begin{array}{ll}
G_\sigma ^{11} & G_\sigma ^{12} \\ 
G_\sigma ^{21} & G_\sigma ^{22}
\end{array}
\right). 
\ee
The function $\zeta(z)$ is defined by the planar series,
$\zeta(z)=\sum_{n=1}^\infty w_n z^{n-1}$ , in which $w_n$ are symmetry 
coefficients given by
\be
w_n=\frac{(-1)^{n+1}(2n-2)!}{n!(n-1)!M^{2n-1}},
\ee
Evaluating the sum we obtain the closed expression
\be
z^2\zeta(z^2)=\frac 12\left( \sqrt{4z^2+M^2}-M\right). 
\label{h}
\ee
Using (\ref{h}) we may write Eq. (\ref{se}) in the form
\be
\hat{\Sigma} _\sigma \hat{P}_\sigma \hat{\Sigma} _\sigma 
+M\hat{\Sigma}_\sigma =\hat{P}_\sigma.
\label{se2} 
\ee
To proceed, we introduce the variables
\be
\theta _{1\sigma }=\frac 1N{\rm Tr}(G_\sigma ^{12})=\sigma \theta _1
\ee
and
\be
\theta _{2\sigma }=\frac 1N{\rm Tr}(G_\sigma ^{21})=\sigma \theta _2,
\ee
which implies the following simple form for the matrix $\hat{P}_\sigma$
\be
\hat{P}_\sigma =N\sigma \left( 
\begin{array}{ll}
0 & \theta _1 \\ 
\theta _2 & 0
\end{array}
\right) \otimes 1_M.
\ee
Using the ansatz
\be
\hat{\Sigma} _\sigma =\sigma \left( 
\begin{array}{ll}
0 & \beta _1 \\ 
\beta _2 & 0
\end{array}
\right) \otimes 1_M
\label{an}
\ee
in Eq. (\ref{se2}), we obtain the following
coupled system
\be
\left\{
\begin{array}{cc}
\beta _1^2\theta _2+2\beta _1=\theta _1 \\
\beta _2^2\theta _1+2\beta _2=\theta _2.
\end{array}
\right.
\label{eq}
\ee
\par
At this stage we need to further specify the nature
of the contacts between the cavity and the leads. 
We assume the presence of asymmetric barriers with
$N$ equivalent transmitting channels.
The adequate choices for the matrices $T$, $L$ and $R$ 
are therefore given by
\be
T=\left( 
\begin{array}{ll}
i\sqrt{\Gamma _1} & 0 \\ 
0 & i\sqrt{\Gamma _2}
\end{array}
\right) \otimes 1_N=L
\ee
and
\be
R=\left( 
\begin{array}{ll}
\sqrt{1-\Gamma _1} & 0 \\ 
0 & \sqrt{1-\Gamma _2}
\end{array}
\right) \otimes 1_N=\bar{S},
\ee
in which $\Gamma_1$ and $\Gamma_2$ represent the transmission
coefficients of each channel in barriers 1 and 2 respectively.
Combining these expressions with the ansatz (\ref{an}) for
the self-energy matrix, we may calculate explicitly the full
Green's function, from which (after using Eq. (\ref{Proj}))
we obtain
\begin{eqnarray}
\nonumber
\theta _1 &=&\frac{\alpha \Gamma _2+(1-\Gamma _2)\beta _1}{1-\alpha \beta
_2\Gamma _2-\beta _1\beta _2(1-\Gamma _2)}+ \\
&&\frac{(1-\Gamma _1)\beta _1}{1-\alpha \beta _1\Gamma _1-\beta _1\beta
_2(1-\Gamma _1)}
\label{eq1}
\end{eqnarray}
and
\begin{eqnarray}
\nonumber
\theta _2 &=&\frac{\alpha \Gamma _1+(1-\Gamma _1)\beta _2}{1-\alpha \beta
_1\Gamma _1-\beta _1\beta _2(1-\Gamma _1)}+ \\
&&\frac{(1-\Gamma _2)\beta _2}{1-\alpha \beta _2\Gamma _2-\beta _1\beta
_2(1-\Gamma _2)}.
\label{eq2}
\end{eqnarray}
Combining (\ref{eq1}) and (\ref{eq2}) with (\ref{eq}), we obtain the following
non-linear algebraic system
\begin{eqnarray}
\nonumber
\alpha\left( 1-\Gamma_{1}\right)\Gamma_{2}\beta_{1}\beta_{2}^{3}+&&\\
\nonumber
\left[ \left(\alpha^{2}\Gamma_{1}\Gamma_{2}+\left(2\Gamma_{1}-1\right)\Gamma_{2}-\Gamma_{1}\right) \beta_{1}-\alpha\left(1+\Gamma_{1}\right) \Gamma_{2}\right]\beta_{2}^{2} +&&\\
\nonumber
\left[ \left( \Gamma_{1}-2\Gamma_{1}\Gamma_{2}\right) \alpha\beta_{1}+\alpha^{2}\Gamma_{1}\Gamma_{2}+\Gamma_{1}+\Gamma_{2}\right]\beta_{2}
-\alpha\Gamma_{1} =0,&&\\
\label{4grau1}
\end{eqnarray}
and 
\begin{eqnarray}
\nonumber
\alpha\left( 1-\Gamma_{2}\right)\Gamma_{1}\beta_{2}\beta_{1}^{3}+&&\\
\nonumber
\left[ \left(\alpha^{2}\Gamma_{1}\Gamma_{2}+\left(2\Gamma_{2}-1\right)\Gamma_{1}-\Gamma_{2}\right) \beta_{2}-\alpha\left(1+\Gamma_{2}\right) \Gamma_{1}\right]\beta_{1}^{2} +&&\\
\nonumber
\left[ \left( \Gamma_{2}-2\Gamma_{1}\Gamma_{2}\right) \alpha\beta_{2}+\alpha^{2}\Gamma_{1}\Gamma_{2}+\Gamma_{1}+\Gamma_{2}\right]\beta_{1}
-\alpha\Gamma_{2} =0.&&\\
\label{4grau2}
\end{eqnarray}
Finally, inserting the ansatz, Eq. (\ref{se}), into (\ref{Fs}) yields
\begin{equation}
f_1(z)=\alpha^{2}N \left[ 1-\frac{\alpha\Gamma_{2}\beta_{2}}{1-\left( 1-\Gamma_{2}\right)\beta_{1}\beta_{2}}\right]^{-1} ,
\label{f2}
\end{equation}
and
\begin{equation}
f_2(z)=\alpha^{2}N \left[ 1-\frac{\alpha\Gamma_{1}\beta_{1}}{1-\left( 1-\Gamma_{1}\right)\beta_{1}\beta_{2}}\right]^{-1} .
\label{f1}
\end{equation}
Equations (\ref{4grau1}), (\ref{4grau2}), (\ref{f1}) and (\ref{f2})
are the central results of this section. Together with Eq (\ref{dt})
they represent the complete solution of the problem. 
It extends the calculations of Brouwer and Beenakker \cite{bb96} to include
the case of asymmetric barriers (although in their solution
of the symmetric case the channels in the barriers were 
considered non-equivalent). Before
presenting a complete analysis of the generic situation,
we shall discuss below some important 
particular cases.

\subsection{Chaotic Cavity with Symmetric Barriers}

The case of symmetric barriers is described by setting 
$\Gamma_1=\Gamma=\Gamma_2$ and $\beta_1=\beta=\beta_2$ 
in Eqs. (\ref{4grau1}), (\ref{4grau2}), 
(\ref{f1}) and (\ref{f2}), so that we find
\begin{equation}
(\alpha\beta^{2}-2\beta+\alpha)((1-\Gamma)\beta^{2}+
\alpha\Gamma\beta-1)=0,
\label{betasb}
\end{equation} 
and
\begin{equation}
f_1(z)=\alpha^{2}N \left[ 1-\frac{\alpha\Gamma\beta}{1-\left( 1-\Gamma\right)\beta^2}\right]^{-1} =f_2(z).
\label{f12sb}
\end{equation}
The physical root of Eq. (\ref{betasb}) is given by
$\beta=(1-\sqrt{1-\alpha^2})/\alpha=\sqrt{z}-\sqrt{z-1}$,
which yields $f_1(z)=f(z)=f_2(z)$, where
\begin{equation}
\frac{f(z)}{N}=\frac{2\left( \sqrt{z}-\sqrt{z-1}\right) \left( 1-\Gamma\right) +\Gamma/\sqrt{z-1}}{2z\left( \sqrt{z}-\sqrt{z-1}\right) \left( 1-\Gamma\right) +\Gamma\sqrt{z}}.
\label{fsb}
\end{equation} 
Inserting (\ref{fsb}) into (\ref{dt}) we find the average density
\begin{equation}
\rho(\tau)=\frac{N}{\pi}\frac{\Gamma(2-\Gamma)}
{(\Gamma^{2}-4\Gamma\tau+4\tau)\sqrt{\tau(1-\tau)}},
\label{dtsb}
\end{equation}
in agreement with Ref. \onlinecite{amsm02}. The solution obtained by
Brouwer and Beenakker \cite{bb96} is, in fact, a generalization of Eq.
(\ref{dtsb}) for non-equivalent channels. It reads
\be
\rho (\tau )=\sum_{n=1}^N\frac{\Gamma_n(2-\Gamma_n)}
{\pi (\Gamma_n^2-4\Gamma_n\tau +4\tau )
\sqrt{\tau (1-\tau )}},
\ee
which, of course, reproduces (\ref{dtsb}) when $\Gamma_n=\Gamma$
for all $n$.

\subsection{Chaotic Cavity with Two Tunnel Junctions}

This system is described by applying the conditions
$\Gamma_1,\Gamma_2 \ll 1$ in Eqs. (\ref{4grau1}), 
(\ref{4grau2}), (\ref{f1}) and (\ref{f2}). We get
\begin{eqnarray}
\beta_{2}^{2}\alpha\Gamma_{2}-\beta_{2}(\Gamma_{1}+\Gamma_{2})+\alpha\Gamma_{1}&=&0\nonumber\\
\beta_{1}^{2}\alpha\Gamma_{1}-\beta_{1}(\Gamma_{1}+\Gamma_{2})+\alpha\Gamma_{2}&=&0,
\label{betatj}
\end{eqnarray} 
together with
\begin{equation}
f_1(z)=\alpha^{2}N \left[ 1-\frac{\alpha\Gamma_{1}\beta_{1}}{1-\beta_{1}\beta_{2}}\right]^{-1},
\end{equation} 
and
\begin{equation}
f_2(z)=\alpha^{2}N \left[ 1-\frac{\alpha\Gamma_{2}\beta_{2}}{1-\beta_{1}\beta_{2}}\right]^{-1}.
\end{equation} 
From the physical roots of (\ref{betatj}), we obtain $f_1(z)=f(z)=f_2(z)$,
where
\be
f(z)=\frac{N}{z}\left[ 1+\frac{\Gamma_{1}\Gamma_{2}}{(\Gamma_{1}+\Gamma_{2})\sqrt{z(z-\tau_{0})}}\right],
\label{ftj}
\ee 
in which $\tau_{0}=4\Gamma_{1}\Gamma_{2}/(\Gamma_{1}+\Gamma_{2})^2$. Inserting
(\ref{ftj}) into (\ref{dt}) we obtain
\begin{equation}
\rho(\tau)=\frac{N\Gamma_{1}\Gamma_{2}}{\pi(\Gamma_{1}+\Gamma_{2})}\frac{1}{\tau^{3/2}\sqrt{\tau_{0}-\tau)}},
\label{dttj}
\end{equation}
in agreement with Refs. \onlinecite{n95} and \onlinecite{ms}.
\par
Using (\ref{dttj}) we can compute the average value of several physical observables.
Of particular interest are the first three cumulants of the FCS, defined in Eqs.
(\ref{g}), (\ref{p}) and (\ref{r}). We find
\be
\left\langle g\right\rangle _{TJ}=N\frac{\Gamma _1\Gamma _2}{\Gamma
_1+\Gamma _2},
\label{gtj}
\ee
for the average conductance,
\be
\left( \frac{\left\langle p\right\rangle }{\left\langle g\right\rangle }
\right) _{TJ}=\frac{\Gamma _1^2+\Gamma _2^2}{(\Gamma _1+\Gamma _2)^2},
\label{Fanotj}
\ee
for the Fano factor, and
\be
\left( \frac{\left\langle \kappa\right\rangle }{\left\langle p\right\rangle }
\right) _{TJ}=\frac{\Gamma _1^4-2\Gamma _1^3\Gamma _2+6\Gamma _1^2\Gamma
_2^2-2\Gamma _1\Gamma _2^3+\Gamma _2^4}{\left( \Gamma _1+\Gamma _2\right)
^2\left( \Gamma _1^2+\Gamma _2^2\right) },
\label{rptj}
\ee
for the ratio between the average third cumulant and the average shot-noise 
power. The subscript {\it TJ} stands for {\it Tunnel Junction }.
Equations (\ref{gtj}), (\ref{Fanotj}) and (\ref{rptj}) are well known in 
the literature \cite{j96}.

\section {General Analysis and Comparison with Circuit Theory}

In this section we compare the predictions of the above diagrammatic approach with
those of the extended version of circuit theory. The exact agreement that we 
found for various quantities strongly suggests the full equivalence of these 
semiclassical techniques in the entire range of input parameters.

\subsection{Circuit Theory}

Circuit theory was invented by Nazarov \cite{n95} and represents a very powerful tool
for computing ensemble averages of quantum chaotic systems in the semiclassical
regime. It consists of a finite element approach in which the spatial support
of the system is partitioned into a network, containing edges (or connectors) 
and vertices (nodes or terminals). In its simplest version, there are only
two terminals and the system is subject to a fictitious pseudo-potential, 
$\Phi$, with fixed values at the terminals and unknown values at the internal
nodes. The basic principles are: 1) a general law of pseudo-current-voltage ($I-V$)
characteristics for a connector $(i,j)$ subject to a pseudo-potential drop
$\Delta\Phi_{ij}$
\be
I_{ij}(\Delta\Phi_{ij})=\int_0^1d\tau \frac{\tau \rho_{ij} (\tau )
\sin(\Delta\Phi_{ij})}
{1-\tau \sin ^2(\Delta\Phi_{ij}/2)},
\label{Iij}
\ee
where $I_{ij}(\Delta\Phi_{ij})$ is a pseudo-current and $\rho_{ij} (\tau )
=\langle {\rm Tr} \delta(\tau-t_{ij}^{}t_{ij}^{\dagger})\rangle$
is the average transmission eigenvalue density of the connector. 
We defined $t_{ij}$ as the sample specific
transmission matrix of the connector. 2) a generalized Kirchhoff
law for pseudo-current conservation at each node.
\par
The power of this approach depends crucially on the appropriate choice
of the connectors, which in turn depends on our ability to calculate
the correspondent pseudo $I-V$ characteristics. This is the point where
the extended approach, put forward in Ref. \onlinecite{amsm02}, 
differs from Nazarov's
original formulation. In Ref. \onlinecite{amsm02} the 
function $I_{ij}(\Delta\Phi_{ij})$
was calculated directly from the supersymmetric non-linear $\sigma$-model.
The advantage of this procedure is the possibility to take full account
of quantum effects in the description, whenever necessary, such as
near barriers and interfaces, without the need to introduce additional 
assumptions for performing the ensemble averages.

\par
As an example, consider the system studied in previous section, i.e. 
a double barrier chaotic billiard with 
$N$ equivalent channels at each contact. In the extended circuit theory, 
we end up with only two equations \cite{amsm02}
\begin{equation}
I(\phi)=I_{1}(\phi-\theta)=I_{2}(\theta),
\label{current}
\end{equation}
where
\be
I_j(\phi)=\frac{N\Gamma_{j}\sin(\phi)}{1-\Gamma_j\sin^2(\phi/2)}
\ee
is the pseudo $I-V$ characteristics of barrier $j$, interpreted as
a connector. We remark that these equations contain all quantum information
that remain relevant in the semiclassical regime after ensemble averaging.
They are therefore a direct consequence of the maximum entropy principle.
In Ref. \onlinecite{ms} it was found to be convenient to introduce
the following modified pseudo-current
\be
K(x)=\frac{i}{2}I(-2ix),
\label{KI}
\ee
which yields the conservation law
\be
K(x)=K_1(x-y)=K_2(y),
\label{Klaw}
\ee
where
\be
K_j(x)=\frac{N}2\left[ \tanh(x+\frac{1}{2} \alpha_j)+\tanh(x-\frac{1}{2}\alpha_j)\right].
\label{Kbar}
\ee
in which we introduced the constants, $\alpha_j$, via the relation $\Gamma_j={\rm sech}^2(\alpha_j/2)$. Inserting (\ref{Kbar}) into (\ref{Klaw}) yields 
\begin{eqnarray}
\nonumber
\tanh(x-y+\frac12\alpha_1)+\tanh(x-y-\frac12\alpha_1)= && \\
\tanh(y+\frac12\alpha_2)+\tanh(y-\frac12\alpha_2),
\end{eqnarray}
which after using the trigonometric identity
\be
\tanh(x-y)=\frac{\tanh x-\tanh y}{1-\tanh x \tanh y},
\label{id}
\ee
yields the following polynomial equation for the variable
$\xi=\tanh y$
\begin{eqnarray}
\nonumber
\lbrack \Gamma_1(1-\Gamma_2)\tanh x]\xi ^4-(3\Gamma_1\Gamma_2\tanh x)\xi ^2+ &&\\
\nonumber
\lbrack (\Gamma_1\Gamma_2+\Gamma_2-\Gamma_1)\tanh ^2x+2\Gamma_1\Gamma_2-\Gamma_1-\Gamma_2]\xi ^3+ &&\\
\nonumber
\lbrack (\Gamma_1\Gamma_2+\Gamma_1-\Gamma_2)\tanh ^2x+\Gamma_1+\Gamma_2]\xi &&\\
-\Gamma_1\tanh x =0. 
\label{circuit}
\end{eqnarray}
This equation must be supplemented by
\be
K(x)=\frac{N\Gamma_2\xi}{1-(1-\Gamma_2)\xi^2}.
\label{Kfull}
\ee
Equations (\ref{circuit}) and (\ref{Kfull}) were first presented in 
Ref. \onlinecite{ms} and are
the circuit theory equations for the double barrier chaotic billiard. In the
next subsection we shall compare them with Eqs. (\ref{4grau1}), 
(\ref{4grau2}), (\ref{f1}) and (\ref{f2}), obtained from the
diagrammatic method.
\par

\subsection{Comparison between Circuit Theory and the Diagrammatic Approach}

In order to facilitate comparison, let us first establish a direct relation
between the generating function $f_j(z)={\rm Tr} F_j(z)$ of the diagrammatic
technique and the modified pseudo-current $K(x)$ of circuit theory. From (\ref{F1})
we obtain
\be
F_1(z)=\sum_{n=0}^\infty \left\langle C_1\left( S^{\dagger }C_2SC_1\right)
^n\right\rangle \frac 1{z^{n+1}}
\ee
and
\be
F_2(z)=\sum_{n=0}^\infty \left\langle C_2\left( SC_1S^{\dagger }C_2\right)
^n\right\rangle \frac 1{z^{n+1}}.
\ee
Inserting (\ref{Smatrix}) and (\ref{C1C2}) into the above equations we get
\be
F_1(z)=\sum_{n=0}^\infty \frac 1{z^{n+1}}\left( 
\begin{array}{ll}
0 & 0 \\ 
0 & (t^{\dagger }t)^n
\end{array}
\right), 
\ee
and
\[
F_2(z)=\sum_{n=0}^\infty \frac 1{z^{n+1}}\left( 
\begin{array}{ll}
(tt^{\dagger })^n & 0 \\ 
0 & 0
\end{array}
\right). 
\]
Performing the trace, we obtain $f_1(z)=f(z)=f_2(z)$, where
\be
z^2f(z)=Nz+h(1/z),
\label{fz}
\ee
and we introduced the function $h(z)$, defined as
\be
h(z)=\left\langle {\rm Tr}\left( \frac{tt^{\dagger }}{1-ztt^{\dagger }}\right)
\right\rangle 
=\int_0^1d\tau \frac{\tau \rho (\tau )}{1-z\tau }.
\label{hz}
\ee
From (\ref{Iij}), one realizes that 
\be
I(\phi )= h(\sin ^2(\phi/2) )\sin \phi,
\ee
which when combined with (\ref{fz}) yields
\begin{equation}
f(z)=\frac{1}{z}\left( N+\frac{I(\phi)}{2\sqrt{z-1}}\right)
\bigg|_{\sin(\phi/2)=1/\sqrt{z}} .
\end{equation} 
Using the relation between $K(x)$ and $I(\phi)$, Eq. (\ref{KI}),
we may rewrite the above equation as
\be
f(z)=\frac 1z \bigg(N-K(x)\tanh x\bigg)\bigg |_{\sinh x=1/\sqrt{-z}}.
\label{fK}
\ee
Let us consider two simple applications of Eq. (\ref{fK}).
For a chaotic cavity with symmetric barriers, Eqs. 
(\ref{circuit}) and (\ref{Kfull}) yield
\be
K(x)=\frac{N\Gamma\sinh x}{2-\Gamma+\Gamma\cosh x}.
\label{Ksym}
\ee
Inserting (\ref{Ksym}) into (\ref{fK}) we find
\be
f(z)=\frac N{\sqrt{z}}\left( \frac{(2-\Gamma )\sqrt{z-1}+\Gamma \sqrt{z}}
{\sqrt{z-1}\left[ (2-\Gamma )\sqrt{z}+\Gamma \sqrt{z-1}\right] }\right), 
\ee
which can be easily shown to agree with (\ref{fsb}).
For a chaotic cavity with two tunnel junctions, we obtain
from Eqs. (\ref{circuit}) and (\ref{Kfull}) the following expression
\be
K(x)=\frac{N\Gamma_1\Gamma_2\sinh 2x}{2\sqrt{\Gamma_1^2+\Gamma_2^2+
2\Gamma_1\Gamma_2\cosh 2x}}.
\label{Ktun}
\ee
Inserting (\ref{Ktun}) into (\ref{fK}) we obtain
\be
f(z)=\frac{N}{z}\left[ 1+\frac{\Gamma_{1}\Gamma_{2}}{(\Gamma_{1}+\Gamma_{2})\sqrt{z(z-\tau_{0})}}\right],
\ee 
in agreement with Eq. (\ref{ftj}).
\par
We can also invert Eq. (\ref{fK}) to obtain $K(x)$ from the
diagrammatic approach
\begin{equation}
K(x)=\coth(x)\bigg( N-zf(z)\bigg)\bigg|_{z=-\sinh^{-2}(x)} .
\label{KF}
\end{equation}
In particular, we can calculate the average of the cumulants of
the FCS using the formula
\begin{equation}
h_{k+1}\equiv\left\langle {\rm Tr}[(tt^{\dagger})^{k+1}]\right\rangle =\frac{(-1)^{k}2^{k}}{k!}\frac{d^{k}H(x)}{d(\cosh(2x))^{k}}\bigg|_{x=0},
\label{cumulant}
\end{equation} 
where $H(x)\equiv 2K(x)/\sinh(2x)$. The first three cumulants are then
given by
\begin{equation}
\left\langle g\right\rangle =h_{1},~~~ \left\langle p\right\rangle =h_{1}-h_{2},
~~~ \left\langle \kappa\right\rangle =h_{1}-3h_{2}+2h_{3}.
\label{gpr}
\end{equation}
We remark that the function $H(x)$ can also be obtained directly from $f(z)$ 
via the relation
\begin{equation}
H(x)=z\bigg(zf(z)-N\bigg)\bigg|_{z=-\sinh^{-2}(x)} .
\label{H}
\end{equation} 
Motivated by strong numerical evidences \cite{ms}, our basic conjecture is that 
one obtains the {\it same} function $H(x)$
either from Eqs. (\ref{4grau1}), (\ref{4grau2}), (\ref{f1}) and (\ref{f2})
or from Eqs. (\ref{circuit}) and (\ref{Kfull}). 
Although explicit analytic evaluation of $H(x)$ in both approaches, for general
$\Gamma_1$ and $\Gamma_2$, is too cumbersome (albeit possible in principle), 
we can still make some analytical progress in verifying our conjecture  
by expanding $H(x)$ 
in powers of $x$, and evaluating the coefficients as explicit functions of 
$\Gamma_1$ and $\Gamma_2$.
This procedure yields, {\it in both approaches}, the following closed expressions
\be
\left\langle g\right\rangle =N\frac{\Gamma _1\Gamma _2}{\Gamma _1+\Gamma _2}
=\left\langle g\right\rangle _{TJ}
\label{gfull}
\ee
for the conductance,
\be
\frac{\left\langle p\right\rangle }{\left\langle g\right\rangle }=\frac{
\Gamma _1+\Gamma _2-\Gamma _1\Gamma _2}{\Gamma _1+\Gamma _2}\left( \frac{
\left\langle p\right\rangle }{\left\langle g\right\rangle }\right) _{TJ}
\label{Fanofull}
\ee
for the Fano factor, and
\be
\frac{\left\langle \kappa\right\rangle }{\left\langle p\right\rangle }=\frac{
\Gamma _1+\Gamma _2-2\Gamma _1\Gamma _2}{\Gamma _1+\Gamma _2}\left( \frac{
\left\langle \kappa\right\rangle }{\left\langle p\right\rangle }\right) _{TJ}
\label{rpfull}
\ee
for the ratio between the average third cumulant, $\langle \kappa \rangle$, and
the average shot-noise power, $\langle p \rangle$. We also found agreement for
the fourth cumulant. Note that for tunnel junctions, 
we have $\Gamma_1,\Gamma_2 \ll 1$, and (\ref{gfull}),(\ref{Fanofull})
and (\ref{rpfull}) reduce to (\ref{gtj}), (\ref{Fanotj}) and (\ref{rptj}),
as expected. The above expressions together with the forth cumulant 
are in complete agreement with the semiclassical cascade
approach presented in Ref. \onlinecite{nsp02}. They represent
strong analytical evidences for the mathematical equivalence between 
Eqs. (\ref{4grau1}), (\ref{4grau2}), (\ref{f1}) and (\ref{f2}) of the 
diagrammatic technique
and Eqs. (\ref{circuit}) and (\ref{Kfull}) of circuit 
theory, in the description of the asymptotic semiclassical domain
of the Poisson kernel distribution. 

\section{Summary and Perspectives}

We presented a detailed comparison between two well known 
semiclassical approaches, the diagrammatic analysis of the unitary group 
and circuit theory, in the description of quantum transport through 
double-barrier chaotic billiards. The problem was reduced to a comparison
between a pair of coupled non-linear algebraic equations (diagrammatic technique)
and a polynomial equation of fourth order (circuit theory). Exact agreement
was found for a variety of quantities, such as the first four average cumulants 
of the full counting statistics and the average transmission eigenvalue density
for symmetric barriers and tunnel junctions. 
The complete equivalence of these approaches is a non-trivial result,
because the semiclassical concatenation principle, used to derive circuit
theory equations, has no obvious counterpart in the diagrammatic method and
leads to a substantial algebraic simplification of the whole problem.
It is quite amusing to observe that circuit theory turns out to yield a 
irreducible description that intuitively satisfies the ``Occam's razor''
principle of descriptive simplicity \cite{s00}.
\par
An interesting consequence of our result would be the extension of circuit
theory to deal with quantum corrections, which can be systematically
treated in both the supersymmetric non-linear $\sigma$-model and in the
diagrammatic technique. 
From a broader perspective, we expect our result to help establishing a
direct connection between several recent independent developments of both 
circuit theory and the diagrammatic $S$-matrix approach, such as those in 
magnetoelectronics \cite{bthb03} with obvious 
technical advantages.

\begin{acknowledgments}
This work was partially supported by CNPq and FACEPE (Brazilian Agencies).
\end{acknowledgments}


\begin{thebibliography}{}
\bibitem{comment}
We have $\beta=1$ for systems with time-reversal (TR) and spin-rotation (SR)
symmetry, $\beta=2$ for systems in the absence of both TR and SR symmetry,
and $\beta=4$ for systems with broken SR symmetry but in the presence of
TR symmetry.
\end{thebibliography}
\end{document}